\documentclass[12pt]{article} 
\usepackage{amssymb,theorem}

\newenvironment{pf}{\noindent{\bf Proof:}}{\newline\kbn}

\newtheorem{satz}{Theorem}
\newtheorem{folge}[satz]{Corollary}
\newtheorem{lemma}[satz]{Lemma}
\newtheorem{prop}[satz]{Proposition}

\newcommand{\Einsop}{\leavevmode{\rm 1\mkern  -4.4mu l}}
\newcommand{\Seins}{{\mathsf{S}^1}}
\newcommand{\cg}{{C}}
\newcommand{\ham}{{H}}

\newcommand{\lok}[1]{{\mathcal #1}}
\newcommand{\Hilb}[1]{{\mathcal #1}}
\newcommand{\Name}[1]{{\sc #1}} 
\newcommand{\dopp}[1]{{\mathbb #1}}
\newcommand{\symb}[1]{{\mathbf #1}}

\newcommand{\qed}{}
\newcommand{\kbn}{$\square$}

\begin{document}

\title{Conformal Transformations as Observables}
\author{S\o{}ren K\"oster\\ Inst.\ f.\ Theor.\ Physik\\ Universit\"at G\"ottingen\\ 37073 G\"ottingen \\ Germany}
\date{}

\maketitle
\begin{abstract} 
$C$ denotes either the conformal group in $3+1$ dimensions,
$PSO(4,2)$, or in one chiral dimension, $PSL(2,\dopp{R})$. Let $U$ 
be a unitary, strongly continuous representation of $C$
satisfying the spectrum condition and 
inducing, by its
adjoint action, automorphisms of a \Name{v.\-Neu\-mann}
algebra $\lok{A}$. We 
construct the unique inner representation $U^\lok{A}$ of the universal
covering group of $C$ implementing these automorphisms. $U^\lok{A}$ satisfies
the spectrum condition and acts trivially on any $U$-invariant vector.

This means in particular: Conformal transformations of a field theory
having positive energy are weak  limit points of 
local observables. Some immediate implications for chiral subnets are
given. We propose the name
``\Name{Bor\-chers-Su\-ga\-wa\-ra} con\-struc\-tion''. \\[1mm]
\end{abstract}


\section{Introduction}
\label{sec:intro}

Space-time symmetries are of paramount importance to relativistic
quantum field theory. Intuitively we expect such coordinate
transformations to be connected to observables. Time translations, for
example, should be observable due to their connection with the energy
operator. If we have a stress energy tensor in the theory, as it is
often the case in models, the energy operator itself is given
as an integral of this local quantum field. Yet, the implementation of
covariance may be given in abstract terms or may stem
from a larger theory into which the theory of interest is embedded (eg
\Name{Coset} models), and it is not always manifest how covariance
may be implemented by observables of the subtheory.

More specifically, as a fact of life any observation is of finite
extension in space and time  and thus we regard the {\em local}
observables as {\em the} constituting objects in quantum field theory. 
For this reason we shall work with the
\Name{v.\-Neu\-mann} algebra $\lok{A}$ which is generated by all local
observables, in 
accordance with the principles given by \Name{Haag, Kastler}
\cite{HK64} and others (cf. \cite{rH92}). 
Thereby our setting also includes quantum field theories  which are
not necessarily 
described completely by covariant quantum {\em fields}, and which might not
possess a stress energy tensor. In fact, the main result is an abstract
statement about  \Name{v.\-Neu\-mann} algebras, without reference to the local
structure of a quantum field theory.

 We consider
representations of such theories which admit a unitary implementation
of covariance and the task thus amounts to a search for observable, unitary,
implementing operators. Quite obviously these operators can not be {\em local}
observables, since  locality implies that adjoint action of these
operators 
is trivial on algebras which are associated with causally disconnected
regions. On the
other hand we believe any observation has to 
be local in nature and we conclude: space-time transformations should
be non-local limits of local observables. We take this as
definition of {\em global observables}.

The problem of identifying space-time symmetry transformations as
global observables is of any  interest only, if the given
representation is reducible. In irreducible representations, such as
the vacuum representation, {\em any} bounded
operator can be represented as a weak limit of local operators. The
representations we have in mind are manifestly reducible and the innerness
of implementing operators in the global sense promises to be of some
use in these circumstances. We return to this point in the latter part
of this work.

To our knowledge this problem so far has been dealt with only in the case of
abelian groups of translations satisfying the spectrum condition
(positivity of energy). \Name{Bor\-chers} \cite{hjB66} solved this
problem relying almost entirely on the spectrum condition and using a
deep result on the innerness of norm-continuous connected
automorphism groups  of \Name{v.\-Neu\-mann} algebras \cite{KR67} (Corollary
8). His result is the key building block in our work. 

 In the abelian case there are many inner implementing representations
with different spectral properties. It was a
challenging task to ensure 
existence of an inner implementing representation satisfying the
spectrum condition. 
\Name{Arveson} \cite{wA74} gave a proof for a
one-parameter group, \Name{Bor\-chers} and \Name{Buchholz}
\cite{hjB87} (and
references therein) succeeded in
solving this problem in general.

In this respect the situation for an inner implementing representation
of $\cg{}$ is different. Because  $\cg{}$ is identical
with its commutator subgroup,
the result of our construction is unique  and validity of the
spectrum condition follows. We show as well that $U$-invariant vectors
are left invariant by the action of the inner implementing
representation $U^{\lok{A}}$. Another result is the proof of
complete reducibility of $U^{\lok{A}}$ making weak assumptions on the
original representation $U$.

In the course of our argument we will construct an inner implementing
representation  $U^{\lok{A}'}$ for the commutant of the
\Name{v.\-Neu\-mann} algebra $\lok{A}$ as well. We have the following relation:
$ U(g) \,= \,\,U^\lok{A}(g) \,U^{\lok{A}'}(g)\, , \quad \forall
\,g\in \cg{}$.
This equation reminds of the \Name{Coset} construction \cite{GKO86}
involving stress energy tensors of chiral current
algebras, which are given by the
\Name{Su\-ga\-wa\-ra} construction \cite{hS68}. It is not difficult to show
that our result agrees with the 
outcome of integrating the respective stress energy tensors.

 At this
point we stress that, although the 
relation to \Name{Coset} constructions as considered by \Name{Goddard,
  Kent and Olive} \cite{GKO86} motivated this work, our result
is independent of the existence of a stress energy
tensor. We make use of this and connect it to a generalised notion of
\Name{Coset} construction, which we discuss briefly.

We make some immediate remarks relating our construction to chiral
subtheories, general chiral \Name{Coset} theories and conformal
inclusions in the latter part of this work. We think
this gives sufficient evidence  for the relevance of an inner
implementing representation in studies on chiral theories. We hope it
will serve well as a substitute for the \Name{Su\-ga\-wa\-ra} stress energy
tensor in many respects, although there are special features of an
inner implementing representation connected to a stress energy
tensor. On the other hand we believe our construction is somewhat
special to (chiral) conformal field theories as we argue in the
discussion concluding this article, and we know that the deeper part
of it is due to \Name{Bor\-chers}. Summing up these thoughts we consider
the term ``\Name{Bor\-chers}-\Name{Su\-ga\-wa\-ra} construction'' appropriate.

We treat the cases $\cg{}=PSO(4,2)$ and $\cg{}=PSL(2,\dopp{R})$ explicetly,
because detailed results on these groups are readily available. We
expect our results to hold true for all conformal groups since we make
use of typical features of conformal groups only.

\section{Preparations and first remarks}
\label{sec:prep}

We deal with the conformal group in $1+3$ dimensions ($PSO(4,2)$) and
in one chiral dimension ($PSL(2,\dopp{R})$). Since both groups share
all the features required here, the symbol $\cg$ will denote both of
them in the following. For geometrical interpretation and some general
facts on $\cg{}$ we refer to \cite{LM75}\cite{gM77}\cite{BGL93}\cite{FG93}.

We use the symbol $\widetilde{\cg}$ for the universal covering group of
$\cg$ and $\symb{p}$ for the covering projection from
$\widetilde{\cg}$ onto $\cg$. The following subgroups of $\cg$ will occur
frequently: The group of translations, $T$, of special conformal
transformations, $S$, the group of global scaling by a factor
$\lambda\in \dopp{R}_+\setminus\{ 0 \}$, $D$, and the
group of ``conformal time'' translations, $R$, which is generated by
the conformal Hamiltonian. The corresponding subgroups of
$\widetilde{\cg}$ will be denoted by $\widetilde{T}$, $\widetilde{S}$,
$\widetilde{D}$, $\widetilde{R}$. We use parameters on $R$ which make
it naturally isomorphic to $\dopp{R}/2\pi\dopp{Z}$. 

 By these conventions we have $R(2\pi)=id$  and the following relation between
the generator of rotations, $\ham{}$, the generator of ``physical
time'' translations, $P_0$,
and the generator of special conformal transformations in direction of
``physical time'', $K_0$:
\begin{equation}
  \label{eq:l0def}
  2 \, \ham{} = P_0-K_0
\end{equation}

In the following $\Hilb{H}$ always stands for a separable
\Name{Hilbert} space,  $U$ and $\widetilde{U}$ are unitary, strongly continuous
representations of $\cg$, $\widetilde{\cg{}}$ on $\Hilb{H}$,
respectively.  We use the physicists' convention on the abstract
\Name{Lie} algebra of $\cg$ and will not 
distinguish between elements of the abstract 
\Name{Lie} algebra and corresponding {\em selfadjoint} generators of unitary,
strongly continuous representations, since this leads to no
ambiguities. If not stated otherwise $\lok{A}$ 
stands for a \Name{v.\-Neu\-mann} algebra of operators on $\Hilb{H}$,
$\lok{A}'$ for its commutant and $\alpha$, $\alpha'$ for automorphic
actions of $\cg{}$ on $\lok{A}$,  $\lok{A}'$ respectively. We
note that any spatial automorphism of $\lok{A}$, given by the adjoint
action of a unitary operator, induces a spatial
automorphism of $\lok{A}'$ as well.

We prove a lemma on the spectrum condition first. The
result is well known (see eg \cite{GL96}, Lemma B.4) and our proof
is not new, presumably, but to our knowledge not yet accessible in the
literature. The argument is short and straightforward; its second part
is adapted from \cite{gM77}. Afterwards we prove uniqueness of the
inner implementing representation.

\begin{prop}\label{prop:poengcond}
  If any one of the
  operators $\ham{}$, $P_0$, $-K_0$ has positive spectrum, then all three of
  them. In this case we say that $\widetilde{U}$ satisfies the
  spectrum condition. 
\end{prop}

\begin{pf}
Assume $\ham{}$ is positive. Take any vector $\phi$
analytic for the
representation $\widetilde{U}$ (cf. eg \cite{BR77}). We have:
\begin{equation}
  \label{eq:scalpos}
  0 \leq 2 \langle\phi, \widetilde{U}(\widetilde{D}(\lambda))\ham{}
    \widetilde{U}(\widetilde{D}(\lambda))^*\phi\rangle = \lambda^2{}
    \langle\phi, P_0\phi\rangle +
  \lambda^{-2} \langle\phi, -K_0\phi\rangle 
\end{equation}
Multiplying by $\lambda^{\pm 2}$ and taking the appropiate 
limits $\lambda\rightarrow 0, \infty$ we deduce  $\omega_\phi(P_0)\geq 0$ and
$\omega_\phi(-K_0)\geq 0$. Since the analytic vectors for the
representation $\widetilde{U}$ form a core for all generators we may apply
criterion 5.6.21 of \cite{KR83}.

Now assume $P_0$ or $-K_0$ is positive. Special conformal transformations
and translations are conjugate in $\cg{}$: $S(-n)=
R(\pi) T(n) R(-\pi)$. 
Defining $g_t$ as $S(n) R(t) T(n) R(-t)$ this identity
becomes: $\lim_{t \nearrow \pi} g_t = id$. Now we see that the corresponding
holds true in $\widetilde{\cg{}}$, since we know it for $\cg{}$, the
relation is continuous in $n$, and the
covering projection is continuous as well. Because conjugation by a unitary
operator does not change the spectrum, positivity of $P_0$ follows from
positivity of $-K_0$ and vice versa. Positivity of $\ham{}$ follows from
equation \ref{eq:l0def} by criterion 5.6.21 of \cite{KR83} applied as
before while discussing equation \ref{eq:scalpos}.\qed
\end{pf}

\begin{prop}\label{prop:inneruni}
  Assume  $Ad_{U}$ induces an automorphism
  group $\alpha$ on $\lok{A}$. If there exists a representation $U^\lok{A}$ of
  $\widetilde{\cg{}}$ by unitary operators in  $\lok{A}$ implementing
  $\alpha$ by its adjoint action on $\lok{A}$, then 
  this representation is unique.
\end{prop}

\begin{pf} Assume there are two such representations, $U_1^\lok{A}$
and $U_2^\lok{A}$. Then the operators
$U_1^\lok{A}(g)U_2^\lok{A}(g)^*$, $g\in\widetilde{\cg{}}$, implement the
trivial automorphism. For this reason these operators belong to the
centre of $\lok{A}$. Using this fact it is straightforward to show
that the operators  $U_1^\lok{A}(g)U_2^\lok{A}(g)^*$ form a
representation of $\widetilde{\cg{}}$. This representation is abelian and
its kernel contains all elements of the form
$g_1g_2g_1^{-1}g_2^{-1}$. Now these elements generate the whole of
$\widetilde{\cg{}}$ since $\widetilde{\cg{}}$ has a simple \Name{Lie}
algebra. Thereby $U_1^\lok{A}(g)U_2^\lok{A}(g)^*=\Einsop$ $\forall
g\in\widetilde{\cg{}}$.\qed\end{pf}

We call a representation $U^\lok{A}$ in the sense of the proposition
above an {\em inner implementing representation} (corresponding to the
pair $(U,\lok{A})$). We immediately have:

\begin{prop}\label{prop:uastrich} 
 Assume the unique inner
 implementing representation 
 $U^\lok{A}$ to exist. Then $U^{\lok{A}'}\equiv U^\lok{A}{}^* \cdot
 U\circ \symb{p}$ is the unique  inner implementing representation
 corresponding to $(U,\lok{A}')$. If $U^\lok{A}$ is  strongly
 continuous, then so is $U^{\lok{A}'}$. 
\end{prop}

\begin{pf} First we prove innerness of the operators 
$U(g) U^\lok{A}(g)^*$ by recognising that their adjoint action on
$\lok{A}$ implements the trivial automorphism. Making use of this
it is straightforward to show that these operators do in
fact define a representation. The implementation property and
unitarity is trivial.  Uniqueness follows from proposition
\ref{prop:inneruni} directly. Continuity is fulfilled, since we are multiplying
continuous functions.
\qed\end{pf}

\section{Realising the construction}
\label{sec:bosug}

This section contains the derivation of our main result. We depend on
the following statement:
\begin{lemma}\label{th:borchers}
  Let $U$ satisfy the spectrum
  condition and let $Ad_U$ induce an automorphism group $\alpha$ of
  $\lok{A}$. Then there  
  are strongly continuous, unitary, inner implementing  representations
  $T^\lok{A}$, $S^\lok{A}$ for the restrictions of
  $\alpha$ to the one parameter subgroups of translations and special
  conformal transformations, respectively.
\end{lemma}

\begin{pf} This is an application of
\Name{Bor\-chers}' theorem \cite{hjB66} and proposition \ref{prop:poengcond}.\qed\end{pf}

At this point we stress that it is not clear at all
whether these restricted inner implementing groups form a
representation of $\widetilde{\cg{}}$. We will show that the inner
implementing representation may be constructed from any given 
pair $T^\lok{A}$, $S^\lok{A}$.
Translations and special conformal transformations together generate
the whole of $\widetilde{\cg{}}$.\footnote{See eg \cite{BGL93} (proposition
1.6), for $PSL(2,\dopp{R})$ use the \Name{Iwasawa} decomposition
\cite{FG93} and do straight forward calculations} The fact that there
are sufficiently many subgroups satisfying the spectrum condition
seems to be special for conformal groups.

\begin{satz}[main theorem]\label{hauptsatz}
Let $U$ be a unitary, strongly continuous representation of
  $\cg{}$ on a
  separable \Name{Hilbert} space $\Hilb{H}$ satisfying the spectrum
  condition, $\lok{A}$ a \Name{v.\-Neu\-mann} algebra of bounded operators
  on $\Hilb{H}$. Assume that the adjoint actions of $U$ on $\lok{A}$,
  $\lok{A}'$ 
  define groups $\alpha$, $\alpha'$ of automorphisms of $\lok{A}$,
  $\lok{A}'$, respectively. 

Then there exist unique unitary, strongly continuous, inner
implementing representations $U^\lok{A}$, $U^{\lok{A}'}\equiv
U^\lok{A}{}^*\cdot U\circ \symb{p}$ of $\widetilde{\cg}$.
\end{satz}

\begin{pf}
We follow arguments given in \cite{BGL95} and look at the unitary group $G$
generated algebraicly by the operators $T^\lok{A}$, $S^\lok{A}$ of
lemma \ref{th:borchers}. We define for any non trivial relation
$\prod_i T^\lok{A}(x_i) S^\lok{A}(n_i) = \Einsop$ a corresponding
element: $g_\pi := \prod_i T(x_i) S(n_i)$. By the implementation
property  of lemma  \ref{th:borchers} we have $\alpha_{g_\pi} (A) = A$
for all $A\in\lok{A}$. The elements $g\in \cg$ having trivial
automorphic action $\alpha_g$ on $\lok{A}$ form a normal subgroup. But
since $\cg$ has trivial centre (\cite{gM77}, direct calculation on
$PSL(2,\dopp{R})$) and simple \Name{Lie} algebra it is simple as a
group and, therefore, we have $g_\pi=id$.

Thus the mapping $\phi: G \rightarrow \cg$ defined by
$T^\lok{A}(x_i)\mapsto T(x_i)$, $S^\lok{A}(n_i)\mapsto (n_i)$ extends
to a surjective homomorphism as translations and special conformal
transformations generate $\cg$. Now we look at the kernel of $\phi$,
$ker_\phi$, and take arbitrary $V\in ker_\phi$. Then we have $V
A V^* = \alpha_{\phi(V)} (A) = A$ for all $A\in \lok{A}$. This implies
that $ker_\phi$ is a central subgroup of $G$. Therefore we have the
following exact sequence, which defines a central extension of $\cg$
by $ker_\phi$:
\begin{displaymath}
  id \quad \longrightarrow \quad ker_\phi \quad \longrightarrow \quad G \quad
  \longrightarrow \quad \cg \quad \longrightarrow \quad id
\end{displaymath}

Now we know that $G$ is a ``weak \Name{Lie} extension'' of $\cg$ in
the sense of \cite{BGL95}. $\cg$ has a simple \Name{Lie} algebra and
because of this it is identical with its commutator subgroup and has vanishing
second cohomology. By the same argument as for the proof of corollary
1.8 \cite{BGL95} we have: there is a unitary, strongly continuous
representation $U^\lok{A}$ of $\widetilde{\cg}$ such that $\phi\circ
U^\lok{A} = \symb{p}$. In particular $U^\lok{A}$ is inner and
implementing.

The remainder follows by proposition \ref{prop:uastrich}.
\qed
\end{pf}

\noindent {\em Two Remarks:} Since we start with a proper
representation  $U$ of 
$\cg{}$, the cocycles of $U^\lok{A}$, $U^{\lok{A}'}$ as (generalised)
ray representations of $\cg$ have to
be mutually inverse, and common eigenvectors of $\ham{}^\lok{A}$,
$\ham{}^{\lok{A}'}$ have eigenvalues which sum up to integers.

In \cite{sK02v2} an alternative derivation was given for
$PSL(2,\dopp{R})$, which applies to representations $\widetilde{U}$ of
$\widetilde{\cg}$ instead of representations $U$ of $\cg$ as well. An
explicit continuous mapping from $\widetilde{\cg}$ into the group of
unitaries of $\lok{A}$ is given there, which yields implementers of the
automorphic action of $\widetilde{\cg}$ on $\lok{A}$. These
implementers thus form a (generalised) ray representation of
$\widetilde{\cg}$, which can be lifted to a proper representation of
$\widetilde{\cg}$. This approach is complementary to the one used here
and agrees with the one used by \Name{Buchholz et al.} \cite{BDF00} (appendix)
for deriving a representation of the \Name{Poincar\'{e}} group from
modular conjugations of wedge algebras.

\section{Examining the result}
\label{sec:exam}

In this section we derive three features of the inner implementing
respresentations which they inherit from the original representation:
spectrum condition, invariant vectors, complete reducibility. We
consider them in this order.

\begin{folge}\label{cor:poseng}
  Both $U^\lok{A}$ and $U^{\lok{A}'}$
  satisfy the spectrum condition. 
\end{folge}

\begin{pf} The operators
$U^{\lok{A}\vee\lok{A}'}(g,h):=U^\lok{A}(g)U^{\lok{A}'}(h)$ define a
unitary, strongly continuous representation of $\widetilde{\cg{}}\times
\widetilde{\cg{}}$. With respect to $U^{\lok{A}\vee\lok{A}'}$ we have a
dense domain of analytic vectors and we take an arbitrary vector
$\psi$ from it. The result follows now as in the
proof of proposition \ref{prop:poengcond} 
from the inequality
$
  0 \leq
  \langle U^\lok{A}(\widetilde{D}(\lambda))^*\psi,P_0U^\lok{A}(\widetilde{D}(\lambda))^*\psi\rangle =
  \langle\psi,\lambda^2{}P^\lok{A}_0\psi\rangle +
  \langle\psi,P^{\lok{A}'}_0\psi\rangle
$.
\qed\end{pf}

\begin{folge}
Let $\Hilb{H}\ni\Omega$ be a vector left invariant by $U$. Then
$U^\lok{A}$, $U^{\lok{A}'}$ both leave $\Omega$ invariant.   
\end{folge}

\begin{pf} 
Since translations and special conformal transformations generate the
whole of $\widetilde{\cg{}}$ it is sufficient to show invariance of
$\Omega$ for these two subgroups. We consider translations only; the
argument for special conformal transformations is the same. We may
specialise further to translations by $t x$, $x\in \dopp{R}^{d+1}$,
$x^2> 0$, $t\in  \dopp{R}$, since any vector in spacetime may be
represented as difference of two timelike vectors. The generator of
translations in direction $x$ is positive by the spectrum
condition. The following argument applies in the case $\cg =
PSL(2,\dopp{R})$ directly.

Take arbitrary $\psi\in\Hilb{H}$. We have
$\langle\psi,U^\lok{A}(g)\Omega\rangle=\langle\psi,U^{\lok{A}'}(g)^*\Omega\rangle$
by assumption.
Set $f_\psi(t):=\langle\psi,U^\lok{A}(\widetilde{T}_x(t))\Omega\rangle$,
$g_\psi(t):=\langle\psi,U^{\lok{A}'}(\widetilde{T}_x(t))^*\Omega\rangle$.
Due to the spectrum condition (corollary \ref{cor:poseng}) $f_\psi$
may be extended to the upper half of the complex plane by means of the
\Name{Laplace} transform (cf. eg \cite{SW64}, chapter 2). This
continuation is analytic in the interior and of at most polynomial
growth for complex arguments. On the real line we have
$|f_\psi|\leq\|\Omega\|\,\|\psi\|$ and due to the theorem of
\Name{Phragmen-Lindel\"{o}f} \cite{eT39} (section 5.62) this bound
holds true for the continuation of $f_\psi$ as well. 

The same line of argument works for $g_\psi$ with respect to the lower
half of the complex plane. Since $f_\psi$ and $g_\psi$ coincide on the
real line both are restrictions of an entire function (reflection
principle). This entire function is bounded by
$\|\Omega\|\,\|\psi\|$, and due to \Name{Liouville}'s theorem
it is constant. Since the vectors $U^\lok{A}(\widetilde{T}_x(t))\Omega$,
$U^{\lok{A}'}(\widetilde{T}_x(t))^*\Omega$ are determined by the scalar
products $f_\psi(t)$ and $g_\psi(t)$, $\psi\in\Hilb{H}$, invariance
follows by taking $t=0$.
\qed\end{pf} 

For the next corollary we prepare ourselves by a lemma and a
comment. In the corollary the representation $U$ is assumed completely 
reducible with finite multiplicities. Although this is a pretty strong
assumption in group theoretical terms, we consider this a rather natural
assumption from the quantum field theoretical point of view. In this
context it is somewhat weaker than a common nuclearity condition
\cite{BGL93}. Nuclearity is desirable for quantum field theories
 and in our setting it corresponds to demanding
the $\ham{}$ eigenspaces to be finite dimensional with
degeneracies growing at most exponentially. Typical (integrable)
chiral models such as current algebras exhibit this behaviour
(cf. eg \cite{FG93}).  This implies our
assumption as the following lemma clarifies.

$\widetilde{R}(2\pi)$ generates an infinite cyclic group contained
in the centre of $\widetilde{\cg}$.\footnote{For $\cg = PSO(4,2)$
  there is an additional $\dopp{Z}_2$ contained in the centre
  \cite{gM77}, but this poses no problem.} The following lemma shows
that complete reduciblity of a representation $\widetilde{U}$ of
$\widetilde{\cg}$ satisfying the spectrum condition is equivalent to requiring 
the representation space to have a decomposition
into a direct sum of eigenspaces of $\widetilde{R}(2\pi)$. Due to the
infinite order of the central subgroup generated by
$\widetilde{R}(2\pi)$ this is not obvious. 

\begin{lemma}\label{lem:disspec}
  Assume the spectrum
  of $\widetilde{U}(\widetilde{R}(2\pi))$ to be pure point. Then the
  spectrum of 
  $\ham{}$ is pure point and $\widetilde{U}$ is completely reducible into a direct
  sum of irreducible representations.
\end{lemma}

\begin{pf}
Let $\Hilb{H}_i$ denote the eigenspace belonging to eigenvalue $e^{i2\pi
  h_i}$. The restriction of $\widetilde{U}(\widetilde{R}(t))e^{-i h_i t}$
to $\Hilb{H}_i$ defines a representation of $\widetilde{U}(1)$. This
representation is completely reducible due to the compactness of
$\widetilde{U}(1)$ (cf. eg \cite{BR77}). This proves the claim on the spectrum
of $\ham{}$.

By the spectrum condition there are vectors of lowest
eigenvalue. By the complete analysis of lowest weights in
unitary representations of $\widetilde{\cg{}}$ \cite{gM77}\cite{dG93}
it is known 
which lowest eigenvalues may occur and that the cyclic representations
generated from these lowest weight vectors are irreducible. Taking
such a lowest weight vector, applying to it the linear 
span of the $\widetilde{U}(g)$, $g\in\widetilde{\cg{}}$, and taking the
completion thus yields an irreducible representation space. We may reduce
 by it because of
unitarity. We iterate this procedure and arrive at the second claim
since $\Hilb{H}$ is separable.
\qed\end{pf}

\begin{folge}
  Assume $U$ to be completely reducible with finite
  multiplicities. Then $U^\lok{A}$ and  $U^{\lok{A}'}$ are completely
  reducible. 
\end{folge}

\begin{pf} Denote the lowest weight vectors by
$\varphi_{(d,i)}$, $i$ being the multiplicity index and $d$ the
eigenvalue of $\ham{}$. For any fixed $d$ the $\varphi_{(d,i)}$ span a
finite dimensional \Name{Hilbert} space. This space is left invariant
by the operators $U^\lok{A}(\widetilde{R}(2\pi))$,
$U^{\lok{A}'}(\widetilde{R}(2\pi))$. Both operators may be diagonalised
on this space simultaneously, the result being a mere relabelling of
the irreducible 
subrepresentations of $U$. Now $U^\lok{A}(\widetilde{R}(2\pi))$,
$U^{\lok{A}'}(\widetilde{R}(2\pi))$ both are diagonal on the irreducible
subspaces generated from the ``new'' lowest weight vectors
$\varphi_{(d,i)}'$  and thus on the whole  of $\Hilb{H}$. Now the claim
follows as in the proof of lemma \ref{lem:disspec}.
\qed\end{pf}
 
\noindent{\em Remark:} Nontrivial unitary representations of
$\widetilde{\cg{}}$ are necessarily infinite dimensional and the
multipicity spaces of $U^\lok{A}$
serve as representation spaces for  $U^{\lok{A}'}$ and vice versa. The irreducible
representations of $U^\lok{A}$ and  $U^{\lok{A}'}$ will, therefore, not have
finite multiplicities in general.

\section{Applications to chiral subnets}
\label{sec:subnet}

In this subsection we gather a few immediate implications of the
\Name{Bor\-chers- Su\-ga\-wa\-ra} construction for chiral subnets.
We denote by $\lok{B}$ a chiral conformal precosheaf in its vacuum
representation satisfying common assumptions and properties as given
in \cite{GL96}. The symbol $I$ stands for proper 
intervals contained in $\Seins$. Although the mapping $I\rightarrow
\lok{B}(I)$ does not define a {\em net} in the proper sense of the
term, we will use this term as we want to 
stress the relation of these models to the concept of local quantum
field theories given usually by nets of local algebras. 

We consider a {\em chiral subnet $\lok{A}$ of $\lok{B}$}. The local
algebras of $\lok{A}$ are contained in the ones of $\lok{B}$ and
$\lok{A}$ satisfies the same  assumptions as $\lok{B}$ except
cyclicity of the vacuum. Properties of local algebras $\lok{A}(I)$
such as weak additivity or factor property can be proved on the basis of
modular invariance of $\lok{A}(I)\subset\lok{B}(I)$ \cite{hjB00} (lemma
VI.1.2.(4.)). The symbol  $\lok{A}$ also denotes the
\Name{v.\-Neu\-mann} algebra generated by all local algebras of the net
$\lok{A}$. Thus all prerequisites for the
\Name{Bor\-chers-Su\-ga\-wa\-ra} construction are at our
disposal. Furthermore we know that for $\lok{A}\subsetneq \lok{B}$ the
projection onto the cylic subspace associated to $\lok{A}$ and the
vacuum $\Omega$ is not $\Einsop$ (modular covariance of $\lok{A}$,
cf. eg \cite{hjB00}) and it is contained in $\lok{A}'$ by the
\Name{Reeh-Schlieder}- theorem. This implies that the representation
of $\lok{A}$ is manifestly reducible and the application of the
\Name{Bor\-chers-Su\-ga\-wa\-ra} construction is not in vain. We collect a few
consequences for any chiral subnet first:
\begin{prop}\label{prop:subnets}
  The inner implementing unitaries $U^\lok{A}(g)\neq\Einsop$ are not
  elements of any local algebra. $\lok{A}$ contains non-trivial
  non-local operators,
  the vacuum is not faithful for $\lok{A}$, and the action of
  $Ad_{U^\lok{A}}$ on the local operators of the net $\lok{A}$ is
  ergodic, if $\lok{A}\neq\dopp{C}\Einsop$. 
\end{prop}

\begin{pf} Suppose for some $g\in\widetilde{\cg{}}$ the unitary
$U^\lok{A}(g)\neq\Einsop$ is contained in a local algebra. By locality and
invariance of the vacuum there is a
local algebra $\lok{B}(I)$ such that all vectors $B\Omega$,
$B\in\lok{B}(I)$, remain unchanged when acted upon by
$U^\lok{A}(g)$. Thus, by the \Name{Reeh-Schlieder} property of $\lok{B}$,
$U^\lok{A}(g)$ has to be trivial and the existence of such operators
is denied. 

The kernel of $U^\lok{A}$ has to be different from $\widetilde{\cg{}}$,
else $\lok{A}$ is left invariant 
pointwise by the covariance automorphisms and therefore must be abelian by
locality. But  local algebras of $\lok{A}$ have to be factors as
elements of a chiral subnet. So $\lok{A}\neq\dopp{C}\Einsop$ requires
the existence of operators $U^\lok{A}(g)\neq\Einsop$. These are not
local operators.

Any fixed point of the action
of $Ad_{U^\lok{A}}$ on a local algebra $\lok{A}(I)$ has to be
contained in its centre due to locality. This centre is trivial since
$\lok{A}(I)$ is a factor. $\Omega$ can not be separating, because we
have: $(U^\lok{A}(g)-\Einsop)\Omega=0$.
\qed\end{pf} 

Now we discuss the relevance of the \Name{Bor\-chers-Su\-ga\-wa\-ra}
construction in studies on chiral subnets and the relation to the
\Name{Su\-ga\-wa\-ra} construction to some extent. To this end we make some
simple considerations on \Name{Coset} theories.

In a large class of chiral conformal models such as free fermions and chiral
current algebras there are explicit constructions for the
transformation operators as observables in terms of local quantum
fields (cf. eg \cite{FST89}). In both cases the construction
yields a representation of the whole \Name{Virasoro} algebra. For
chiral current algebras the construction was given by \Name{Su\-ga\-wa\-ra}
\cite{hS68} up to a numerical factor. This diffeomorphism invariance
is broken in any positive energy representation necessarily; it
remains a $\widetilde{\cg{}}$ symmetry only. 

We have constructed the inner implementation of this remaining
symmetry in a completely model independent way. Results of
\Name{Rehren} \cite{khR00} indicate this inner implementing
representation might play a central role in studies on chiral
subnets. One situation of particular interest arises  if one considers
the set of algebras defined by the local relative commutants
$\lok{C}_I:=\lok{A}(I)'\cap\lok{B}(I)$ of a subnet
$\lok{A}\subset\lok{B}$. If one can show this set to satisfy isotony,
it is in fact a chiral subnet $\lok{C}\subset\lok{B}$
itself\footnote{For chiral current subalgebras isotony for the local
  relative commutants (ie the property $\lok{\cg{}_I}\subset\lok{C}_J$ for
  $I\subset J$) follows from strong additivity
  (``\Name{v.\-Neu\-mann} density'') in positive energy representations
  (cf. \cite{vL97}, corollary 1.3.3).}. We define a {\em \Name{Coset}
  theory $\lok{C}$ 
associated to a subnet $\lok{A}\subset\lok{B}$} to be a chiral subnet
$\lok{C}\subset\lok{B}$ satisfying $\lok{C}(I)\subset\lok{C}_I$. A
simple argument leads to the following lemma:
\begin{lemma}
The maximal \Name{Coset} theory associated to a subnet
$\lok{A}\subset\lok{B}$  is defined by
$\lok{C}^{max}(I):=\{U^\lok{A}(g),
  g\in\widetilde{\cg{}}\}'\cap\lok{B}(I)$. It satisfies
$\lok{C}^{max}(I)=\lok{A}'\cap\lok{B}(I)$ as well.
\end{lemma}

\begin{pf} Obviously this definition yields a subnet
$\lok{C}^{max}\subset\lok{B}$. Since the operators of a local algebra
of $\lok{C}^{max}$ commute with the inner implementation of $\lok{A}$,
we deduce from locality of $\lok{B}$ that $\lok{C}^{max}$  is in fact a
\Name{Coset} theory.

Let $\lok{C}$ be any \Name{Coset} theory, $I, J$ proper intervals
satisfying $I\subset J$ and $I'\cup J =\Seins$. By isotony of
$\lok{C}$, locality and weak additivity for chiral subnets we have: 
$
  \lok{C}(I)\subset (\lok{A}(I')\vee\lok{A}(J))' =
  \lok{A}'\subset \{U^\lok{A}(g),
  g\in\widetilde{\cg{}}\}'
$.
\qed\end{pf}

While the global algebra $\lok{A}$ might be a fairly intractable
object, the representing operators have a lot of well known
features. Therefore the characterisation given above may prove
useful. Certainly $U^{\lok{A}'}$ implements covariance on any
\Name{Coset} theory, but it is not obvious whether  $U^{\lok{A}'}$
itself is contained in the global algebra $\lok{C}^{max}$.

It might happen that a subnet $\lok{A}\subset\lok{B}$ admits no
\Name{Coset} theory at all,
i.e. $\lok{C}^{max}(I)=\dopp{C}\Einsop$. In this case we call
$\lok{A}\subset\lok{B}$ a {\em conformal inclusion}. This term stems
from studies on chiral current algebras. Here we have for both nets
$\lok{A}\subset\lok{B}$ stress energy tensors $\Theta^\lok{A}$,
$\Theta^\lok{B}$. A simple argument shows that their difference
$\Theta^\lok{B}-\Theta^\lok{A}\equiv \Theta^{coset}$ is a stress
energy tensor alike. By 
the \Name{Reeh-Schlieder} theorem and the \Name{L\"uscher-Mack} theorem
\cite{FST89} $\Theta^{coset}$ vanishes iff its central charge
vanishes. Its central charge  is completely determined by the finite
dimensional \Name{Lie} algebras from which the 
current algebras are constructed and by the embedding of the smaller one
into the larger one. Its zeros, characterising the notion of conformal
embeddings for these models, have been 
classified \cite{SW86}\cite{BB87}. The following
proposition shows that our definition covers these as special cases.

\begin{prop}
  Suppose the inner implementing representation of theorem
  \ref{hauptsatz} for a chiral subnet $\lok{A}\subset\lok{B}$
  satisfies $U=U^\lok{A}$. Then $\lok{A}\subset\lok{B}$ is conformal.
\end{prop}

\begin{pf} By assumption we have
$U^{\lok{A}'}=\Einsop$. Since $U^{\lok{A}'}$ implements covariance on
any \Name{Coset} theory, the local algebras of $\lok{C}^{max}$ have to
be trivial by the reasoning given in the proof to proposition
\ref{prop:subnets}.
\qed\end{pf}

While given $\lok{A}$ and $U$ the inner implementing representation
$U^\lok{A}$ is unique, $U^\lok{A}$ does not determine the subnet
$\lok{A}\subset\lok{B}$, as examples of conformal embeddings show. In
general there will be a lot of subnets transforming covariantly under
the action of $U^\lok{A}$ (transformation property) and a lot of
subnets containing the operators of $U^\lok{A}$ as global
observables (generating property). Generically  there will be no simple
relation such as inclusion 
or commutativity etc for any pair $\lok{M}_1$, $\lok{M}_2$ of chiral
subnets having one or both properties. There is, of course, a maximal subnet
transforming covariantly and having the generating property. It is
given by:
$ \lok{A}_{max}(I) := \{U^{\lok{A}'}(g), \,\,
    g\in\widetilde{\cg{}}\}'\cap\lok{B}(I) 
$.
Any subnet $\lok{A}$ having both properties defines a conformal
inclusion $\lok{A}\subset\lok{A}_{max}$. Since studies on conformal
inclusions form an area of research of their own, $\lok{A}_{max}$ should
be a generic object to explore.

\section{Discussion}
\label{sec:discus}

We have presented a construction applying and generalising the result
of \Name{Bor\-chers} \cite{hjB66}. The result coincides with the
corresponding structure in special cases in which there is a stress
energy tensor. In particular it generalises, within its natural
limits, the \Name{Su\-ga\-wa\-ra} construction \cite{hS68}. We have proposed the name
``\Name{Bor\-chers-Su\-ga\-wa\-ra} construction'' because of these relations.
The construction is completely model independent and does not require
existence of a stress energy tensor. We expect special features of an
inner implementing representation connected to a stress energy
tensor. This is subject to work in progress.

It is natural to ask if this construction may be applied to other
space-time symmetry groups. In our view the key tools in our
construction are the following: the original representation satisfies
the spectrum condition for some translation subgroups. There are
sufficiently many of them to generate the whole group and we have an
argument how to derive a representation of the covering group from the
unitary group generated by operators constructed by means of \Name{Borchers}'
theorem. 

We have not examined applicability of our strategy to other cases in
any detail, but we want to comment on the \Name{Poincar\'e} group. Here the
translations usually satisfy the spectrum condition. Unfortunately, so
to say, they form an invariant subgroup and although one is tempted to
generate the group from $PSL(2,\dopp{R})$ subgroups (as eg in
\cite{KW01}) this seems impossible with subgroups satisfying the
spectrum condition. Therefore this most important case is still out of
reach. 


\subsection*{Acknowledgements}
I thank \Name{K.-H.Rehren} (G\"ottingen) for
  many helpful discussions and a critical reading of the
  manuscript. Furthermore I owe the idea of using the argument of
  \cite{BGL95} a referee's comment on an earlier version of this paper\cite{sK02v2}. Financial support from the \Name{Ev. Studienwerk
    Villigst} is gratefully acknowledged.


\end{document}